\documentclass[prodmode,acmjocch]{acmlarge}

\acmVolume{0}
\acmNumber{0}
\acmArticle{0}
\articleSeq{0}
\acmYear{2015}
\acmMonth{0}

\doi{0000001.0000001}

\issn{1234-56789}

\usepackage[ruled]{algorithm2e}
\SetAlFnt{\algofont}
\SetAlCapFnt{\algofont}
\SetAlCapNameFnt{\algofont}
\SetAlCapHSkip{0pt}
\IncMargin{-\parindent}

\title{Corpus COFLA: A research corpus for the computational study of flamenco music}
\author{NADINE KROHER\affil{Universitat Pompeu Fabra}
JOS{\'E}-MIGUEL D{\'I}AZ-B{\'A}{\~N}EZ and JOAQUIN MORA
\affil{Universidad de Sevilla}
EMILIA G{\'O}MEZ\affil{Universitat Pompeu Fabra}
}

\begin{abstract}
Flamenco is a music tradition from Southern Spain which attracts a growing community of enthusiasts around the world. Its unique melodic and rhythmic elements, the typically spontaneous and improvised interpretation and its diversity regarding styles make this still largely undocumented art form a particularly  interesting material for musicological studies. In prior works it has already been demonstrated that research on computational analysis of flamenco music, despite it being a relatively new field, can provide powerful tools for the discovery and diffusion of this genre. In this paper we present {\em corpusCOFLA}, a data framework for the development of such computational tools. The proposed collection of audio recordings and meta-data serves as a pool for creating annotated subsets which can be used in development and evaluation of algorithms for specific music information retrieval tasks. First, we describe the design criteria for the corpus creation and then provide various examples of subsets drawn from the corpus. We showcase possible research applications in the context of computational study of flamenco music and give perspectives regarding further development of the corpus.
\end{abstract}


\acmformat{Nadine Kroher, Jos{\'e} Miguel D{\'i}az-B{\'a}{\~n}ez, Joaquin Mora and Emilia G{\'o}mez. 2015. Corpus COFLA: A research corpus for the Computational study of Flamenco Music.}

\begin{document}

\begin{bottomstuff}
This work is supported by the Junta de Andalucia (COFLA2 \#P12-TIC-1362), the Spanish Ministry of Education (SIGMUS TIN2012-36650) and the PhD fellowship of the Department of Information and Communication Technologies, Universitat Pompeu Fabra.\\
Author's address: N. Kroher, E. G{\'o}mez, Music Technology Group, Universitat Pompeu Fabra, Tanger, 122-144, 08018 Barcelona, Spain; email: nadine.kroher@upf.edu, emilia.gomez@upf.edu; Joaquin Mora, Jos{\'e} Miguel D{\'i}az-B{\'a}{\~n}ez, Escuela Superior de Ingenieros, Universidad de Sevilla, Camino de los Descubrimientos, s/n, 41092 Sevilla Spain; email: jmora@us.es, dbanez@us.es;
\end{bottomstuff}

\maketitle

\section{Introduction}
In the past, research on flamenco music has spanned over a variety of disciplines and an active community of researches has formed. Apart from musicological aspects, studies have focused on lyrics, history, evolution and social aspects, among others. Examples of research on a variety of topics related to flamenco can be found in \cite{INFLA} and \cite{INFLA2}. 

Flamenco is an oral music tradition where songs and musical resources have been passed from generation to generation. Consequently, scores are scarce and given the high degree and complexity of melodic ornamentation of the vocal melody, manual annotations are extremely time-consuming and always involve a certain degree of subjective interpretation. Furthermore, the underlying musical concepts and the evolution of flamenco music are still largely undocumented. This fact poses strong limitations to traditional musicological studies of the genre, and, at the same time, provides the main motivation for developing computational tools for flamenco description and analysis. Related work in this relatively young field has mainly been carried out in the scope of the COFLA \footnote{http://www.cofla-project.com/} research project. Apart from automatic transcription of the singing melody (\cite{MonoTrans}, \cite{AT}), previous research has focused on melodic and rhythmic similarity (\cite{melodicSimilarity}, \cite{similarity1}, \cite{similarity2}), melodic pattern detection (\cite{pattern} and \cite{pattern2}), metrical structure identification (\cite{dekai}), perceived emotion (\cite{emotion}), genre classification (\cite{genre}) and intra- and inter-style classification (\cite{similarity}). A comprehensive overview is provided in \cite{Bridges14}. 

Until now, such studies mostly rely on datasets gathered specifically for a particular task and the included tracks often originate from private collections. In order to provide a generic data framework for such studies, suitable for the development, adjustment and evaluation of computational tools for automatic flamenco description and analysis, we gathered the \textit{corpusCOFLA} database: The research corpus consists of audio recordings and editorial meta-data, carefully selected by experts in the field, and provides a data pool (the {\em universe}) for the creation of subsets and test collections for specific music information retrieval tasks. Or aim is to make the corpus as well as the annotated subsets available for the research community in order to provide reproducibility of research outcomes and facilitate data to researchers who are interested in engaging in research activities in the field. Despite the fact that the collection was gathered for the purpose of aiding computational studies, it can support research dealing with flamenco from a variety of disciplines, such as traditional musicological or semantic analysis. 

\subsection{Flamenco music}
Flamenco is an oral music tradition with roots as diverse as the cultural influences of its area of origin, Andalusia, a province in southern Spain. Over the centuries, the area and, of course, its music have been influenced by a variety of settlements of different cultures. Among them, Jews and Arabs, but mostly Andalusian Gypsies have shaped flamenco music to its form as we know it today. Due to its particular characteristics and importance for the cultural identity of its area of origin, flamenco as an art form was inscribed in the UNESCO List of Intangible Cultural Heritage of Humanity in 2010. In \cite{Bridges14}, flamenco is described as an {\em"eminently individual yet highly structured form of music"}, meaning that although interpretations are largely improvisational, the tradition is characterised by an elaborate organisation of styles and structures. This implicit knowledge sets the basis for spontaneous improvisations in which the artists combine the fixed rhythmic, melodic and harmonic structures of a particular style with a set of individual expressive resources. 

The singing voice, usually accompanied by guitar playing and hand-clapping, represents the central and most expressive element of flamenco music. Consequently, the main focus in the computational study of flamenco music is set on developing algorithms which target the analysis of the singing voice. Its particular characteristics are described in \cite{similarity1}: Strong fluctuations of dynamics and timbre, a large amount of melodic ornamentation and the absence of the singer's formant represent key features. Flamenco singers learn by oral transmission in which they acquire a set of melodies corresponding to different styles as well a number of vocal resources used in spontaneous interpretations. Key melodic characteristics are summarised in \cite{Guerrero10}: The combination of short notes (syllabic parts) and long notes (with or without melisma) in flamenco melodies generate the equilibrium of a well-structured lyric along a phrase. Long notes are often ornamented by melisma, consisting of groups of 3, 5, or up to 7 notes. The number and duration of notes in a melisma depend on the ability of the singer to maintain speed and rhythmic control. Melismas are placed in specific locations of a phrase, such as at its end (as a cadence in the last syllable), or within tonic or post-tonic syllables. Dynamic changes are driven by rhythmic accentuation in the style, although there might be some sudden changes in volume caused by expressive traits. For a complete description of the flamenco singing and its diversity of styles, we refer the reader to \cite{historia} and \cite{gamboa05}.

\subsection{Research corpus design}
While a number of datasets for music information retrieval (MIR) research purposes are available, as for example the \textit{million song dataset} \cite{MSD}, their content is mostly limited to Western commercial music. Only recently, a first approach towards creating research corpora for non-Western music traditions has been carried out in the scope of the \textit{CompMusic}\footnote{http://compmusic.upf.edu/} research project: After establishing general guidelines for design of research corpora for computational music studies \cite{Serra14}, corpora have been created for Indian Art Music \cite{Srini14}, Turkish Makam Music \cite{Uyar14}, Beijing Opera \cite{Caro14} and the Arab-Andalusian music tradition \cite{Sordo14}. Given the absence of scores, the strongly improvisational character, the diversity of styles and the lack of documentation, flamenco music poses a particular challenge for the creation of a research corpus. 

A key paradigm established in \cite{Serra14} is the discrimination between a research corpus and a test collection: A research corpus provides a pool of authentic data representative for the particular genre under study. In contrast, test collections contain manual annotations and represent the ground truth for developing and testing algorithms for specific MIR tasks. In this study, we present a research corpus, consisting of carefully selected audio examples and the corresponding editorial meta-data. This generic framework of flamenco music serves as a pool for the creation of test collections targeting particular tasks.  

We adopt the following criteria for the creation of research corpora proposed in \cite{Serra14}:

\begin{description}
  \item [Purpose] Definition of the related research problems and the applied technologies and approaches.
  \item [Coverage] The corpus should cover a representative sample of the music under study and should reflect its variety regarding musical aspects. 
  \item [Completeness] Refers to the integrity of the meta-data accompanying the audio recordings.
  \item [Quality] The audio quality of the included samples must match certain standards, depending on the target applications.
  \item [Reusability] In order to guarantee reproducibility of related research, the data should be accessible to the research community. 
\end{description}

We furthermore gathered three test collections which target specific MIR applications: We manually annotated ground truth which we provide together with meta-data and a number of automatically extracted audio descriptors. 

The remainder of the paper is structured as follows: We first describe the applied design criteria established in related work and subsequently present the gathered research corpus. Next, we describe three test collections drawn from the corpus and present example application in computational flamenco analysis. Finally, we conclude our work and give perspectives for further development of the corpus.

\section{The flamenco corpus}
\begin{figure}[!ht]
 \centerline{
 \includegraphics[width=6.5cm]{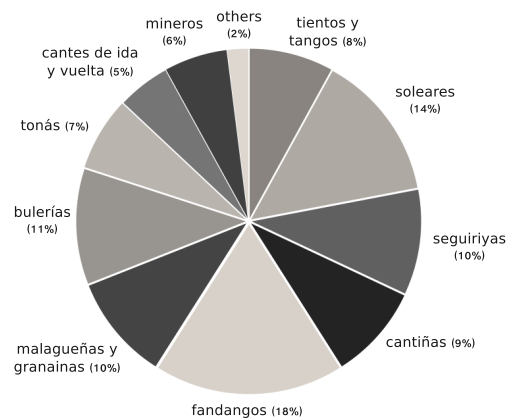}}
 \caption{Distribution of style categories within the corpus.}
 \label{fig:pie}
\end{figure}

\begin{figure}[!ht]
 \centerline{
 \includegraphics[width=15.5cm]{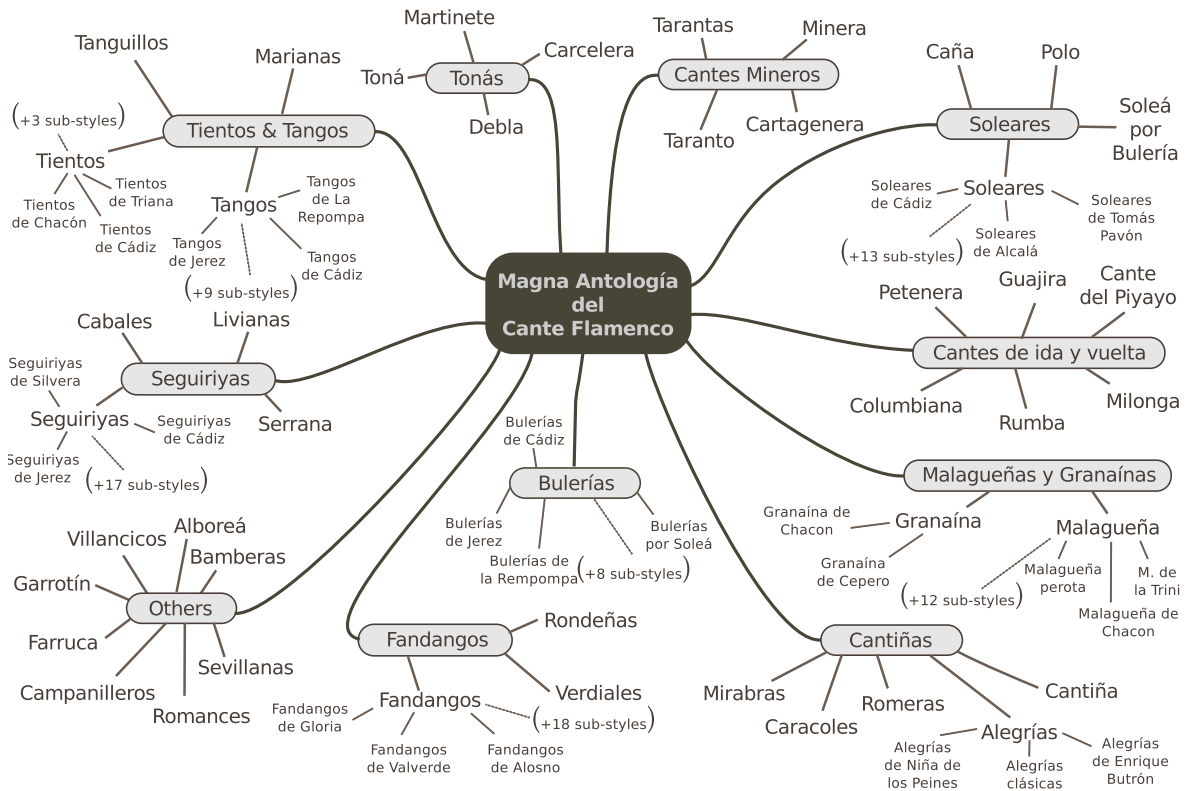}}
 \caption{Styles and sub-styles contained in the {\em magna antilog{\'i}a del cante flamenco}.}
 \label{fig:palos}
\end{figure}

The proposed research corpus, \textit{corpusCOFLA}, consists of more than 1800 audio recordings together with their corresponding editorial meta-data taken from flamenco anthologies. The complete collection encompasses 362 singers and a total of approximately 95 hours of music. We have conducted a statistical analysis of the corpus data with respect to artist and style in order to ensure the representative nature of the collection. Flamenco music is characterised by a hierarchical structure of styles and sub-styles (Figure \ref{fig:palos}) and a systematic classification has not been established so far. In correspondence with flamenco experts we defined ten style families (Figure \ref{fig:pie}) in order to analyse the distribution of recordings in the corpus with respect to style. 

The key characteristics of the corpus can be summarised as follows:

\begin{itemize}
	\item The collection is exhaustive in a sense that it contains all anthologies published on CD during the 20th century and covers all renown recordings of what is considered {\em classical flamenco}.
	\item  These anthologies are known references for the genre and recognised as such by both, music critics and enthusiasts. 
	\item Each anthology is a representative subsample of flamenco music and its diversity, since style- and singer-specific collections are excluded. The diversity of styles as well as their frequency of occurrence in flamenco festivals and concerts is reflected in the corpus (Figure \ref{fig:pie}). 
	\item The size of the corpus is sufficient in order to encompass all singers of significant importance to the 20th century flamenco as well as all essential styles and their variants. 
	\item Given the large time span covered in the collection, the audio quality varies among tracks. Nevertheless, the included anthologies are published by renown record labels and all recordings comply with a minimum standard, which is sufficient for a large variety of audio processing applications.
	\item The included anthologies are commercially available which facilitates the acquisition in the scope of research activities. Furthermore, this fact strengthens the intention of providing suitable ground truth data for computational algorithms which target mainly commercial recordings.
\end{itemize}

An overview of the corpus data is given in Table \ref{tab:statistics} and the design criteria are discussed in detail below.

\begin{table}[t]
\tbl{Corpus and meta-data statistics}{%
\begin{tabular}{|l|c|}
\hline
number of anthologies & 12 \\
total number of CDs & 103 \\
total number of tracks & 1812\\
total number of singers & 362\\
male singers & 81\%\\
female singers & 19\%\\
total duration & approx. 95 hours\\
title annotation existent & 83\% \\
style annotation existent & 94\% \\
\hline
\end{tabular}}
\begin{tabnote}
The corpus comprises more than 1800 tracks with a total duration of approximately 95 hours.
\end{tabnote}
\label{tab:statistics}
\end{table}

\begin{figure}[!ht]
 \centerline{
 \includegraphics[width=6.0cm]{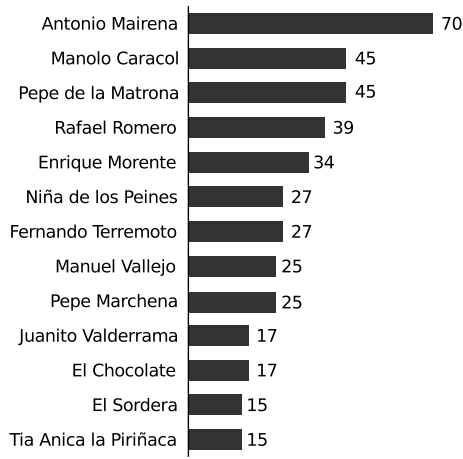}}
 \caption{Distribution by singer: most occurring singers and number of tracks.}
 \label{fig:singers}
\end{figure}

\begin{figure}[!ht]
 \centerline{
 \includegraphics[width=6.5cm]{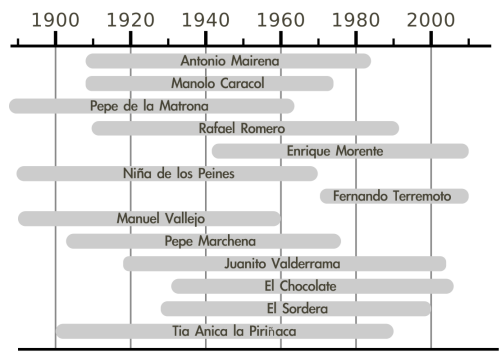}}
 \caption{Timeline displaying the life time of the most frequently occurring singers.}
 \label{fig:timeline}
\end{figure}

\begin{table}[t]
\tbl{Anthologies comprising the research corpus.}{%
\begin{tabular}{| l |c |c |c |c |}
  \hline
  \textbf {Title} & \textbf {Record label} & \textbf {Release} & \textbf {Re-edition} & \textbf{No. CDs} \\
  \hline
  Antolog{\'i}a del Cante Flamenco & Hispavox & 1958 & 1988 & 2 \\
  \hline
  Antolog{\'i}a del Cante Flamenco y Cante Gitano & Columbia & 1965 & 2001 & 2 \\
  \hline
  Archivo del cante flamenco & Vergara & 1968 & 2011 & 4 \\
  \hline
 Magna  Antolog{\'i}a del Cante Flamenco & Hispavox & 1978 & 1992 & 10 \\
  \hline
El Cante Flamenco. Antolog{\'i}a Hist{\'o}rica  & Philips-Universal & 1986 & 2004 & 3 \\
  \hline
 Medio Siglo de Cante Flamenco  & BMG-Ariola & 1987 & 2012 & 4 \\
  \hline
   Antolog{\'i}a de Cantaores flamencos  & EMI-Ode{\'o}n & 1987 & 2003 & 15 \\
  \hline
     Flamencolog{\'i}a, Antolog{\'i}a del Cante Flamenco  & Planet Records & 1993 & 2003 & 7 \\
  \hline
     Antolog{\'i}a del Cante Flamenco  & Orfe{\'o}n-Sony & 1994 &  --- & 4 \\
  \hline
    Historia del flamenco  & Tartessos & 1996 &  --- & 40 \\
  \hline
     100 a{\~n}os de Flamenco  & EMI-Ode{\'o}n & 1997 & --- & 2 \\
  \hline
     Atlas del cante flamenco  & Universal & 2001 & 2011 & 10 \\
  \hline
 \end{tabular}
}
\begin{tabnote}
The corpus is composed of 12 anthologies published under 10 different record labels.
\end{tabnote}
\label{tab:anthologies}
\end{table}

\subsection{Purpose}
We aim to develop methodologies for automatic and computer-assisted description and analysis of flamenco music. Until now, we mainly focus on the singing voice and we consequently target collections of audio recordings where the vocals represent the central musical element. Nevertheless, possible studies of the guitar accompaniment are considered in the corpus design. Our main objectives include the creation of computational tools for large-scale musicological studies and novel computer-assisted methodologies which beyond the traditional score analysis. We furthermore target the automatic description and categorisation of flamenco music to facilitate automatised indexing of music databases and to consequently aid diffusion of the genre. 

We use signal processing techniques to generate quantitative representations of the audio signal in various levels of abstractions, mostly related to melodic and harmonic contents of the analysed track. Even though the target applications are computational studies, the proposed corpus, given its representative nature of the genre, also provides a suitable basis for musicological and inter-disciplinary approaches.

\subsection{Coverage}
In correspondence with experts in the field, we selected a number of flamenco anthologies with the aim of creating a corpus which reflects the diversity of musical concepts inherit to flamenco music. A key goal is to avoid a possible bias towards particular styles, singers, record labels or geographic locations of origin as it is the case for many commercially available collections. We aim for a complete representation of what is considered {\em classical flamenco}, a well-defined concept among flamenco experts comprising a set of established and renowned artists of the 20th century. Recent variants of flamenco are excluded, given their constant evolution and volatile appearance and disappearance. 

Anthologies are a suitable basis for the corpus creation, since they already represent systematically selected collections, aiming to reflect the essence and diversity of the genre and including well established interpreters. Furthermore, the purpose of anthologies is usually for the listener to be able to explore the genre. This feature implies the aim of the editor to gather a representative collection, which coincides with out intentions. The drawback of adopting an existing selection in form of a single anthology is the fact that it is not guaranteed to be unbiased: Apart from personal preferences regarding styles and singers, the editor of an anthology might be restricted to material released by particular record labels. On the other hand, trying to create an unbiased selection for the research purposes mentioned above would require an exhaustive in-depth study of all available recordings on the market as well as the time-consuming process of a number of experts agreeing on a final selection. Even in this case, an implicit bias cannot be excluded and furthermore the reproducibility regarding the acquisition of the audio recordings for other research projects would significantly increase in complexity. 

We therefore decided to create the research corpus based on various anthologies, in order to average possible existing selection tendencies of single collections. In correspondence with experts in the field we selected those anthologies which fulfil the following criteria: The considered collections are selected with the aim of creating a systematic and representative anthology of flamenco music. We exclude miscellaneous collection or those which refer to a single artist, style or geographic location. We  furthermore limit the selection to commercially available collections in digital format and with an audio quality suitable for a variety of computational analysis tasks. The resulting selection of anthologies which comprise \textit{corpusCOFLA} are summarised in Table \ref{tab:anthologies} and the most occurring singers are displayed in Figure \ref{fig:singers} together with their biographical data in Figure \ref{fig:timeline}.

\subsection{Completeness}
Given the absence of scores in flamenco singing, we focus solely on the audio recording and the editorial meta-data. In this case, completeness refers to the integrity of the provided meta-information. For each track, we provide title, singer, style and track duration as annotated by the editorial in a machine-readable text format. This data is incomplete in a sense that some collections do not annotate the style and in some cases the title of the track is missing and replaced by the corresponding style. The statistics summarising the completeness of this data are shown in Table \ref{tab:statistics}. It is worth to mention that editorial style annotations do not follow a strict taxonomy, which furthermore has not yet been established in the context of flamenco music. In order to illustrate the variety of style families, styles and sub-styles and their complex hierarchical structures, the editorial style annotations found in the anthology {\em magna antología del cante flamenco} are displayed in Figure \ref{fig:palos}. An overview of the distribution of style families among the full corpus is given in Figure \ref{fig:pie}. We also detected several ambiguities regarding the artist name. As stated in the last section, future development of the corpus includes the definition and manual annotation of a style taxonomy and the revision and manual correction of editorial meta-data. 

\subsection{Quality}
Since the creation of this research corpus is targeted to the application of signal processing technologies, good audio quality is desired. The final selection contains commercial live and studio recordings by renown record labels, providing acceptable quality for most studies. Nevertheless, given the large time span of production years and the variety of recording circumstances, quality varies among the tracks. We consider this issue when creating test collections for particular tasks, since the required audio quality strongly depends on the target application. 

\subsection{Reusability}
In order to facilitate the use of this research corpus, we ensured that all contained anthologies are commercially available. Given copyright restrictions, the actual audio recordings can not be made accessible in a public web repository. Consequently, all audio tracks are shared on request for research purposes only or can be purchased. We furthermore provide an additional document with editorial information about the collections to simplify their acquisition. The corresponding editorial meta-data is delivered in a machine-readable format allowing automated parsing. All data, including the test collections described in the next section, are publicly available\footnote{http://www.cofla-project.com/corpus.html}. 

\section{Test collections}
As described at the outset, the proposed research corpus provides a representative sample of classical flamenco music and is suitable for explorative studies. For specific music information retrieval problems and the development and evaluation of novel systems and algorithms, we need to create annotated test collections, providing the ground truth for the respective task. In the scope of the {\em corpusCOFLA} project, we gathered three such subsets which can support a number of related applications: The \textit{cante2midi} set contains manual transcriptions of the singing voice melody and the {\em canteFAN} collection contains manual annotations of repeated melodic patterns. The {\em cante100} subset represents a small-scale sample representative of the corpus. It was gathered with a uniform sampling with respect to style families and can be used in the context of inter- and intra-style characterisation as well explorative studies for a variety of tasks. 

The problem of documenting annotated data collections was recently addressed in \cite{Peeters12}, where a systematic description scheme was established with the objective to encourage the re-use of existing collections throughout the community. We adopt this scheme and give the corresponding descriptions in the appendix. 

We provide editorial meta-information for all tracks in machine-readable format and also incorporate this data in the open music encyclopaedia {\em MusicBrainz}\footnote{http://www.musicbrainz.org/}. In addition, we provide a variety of automatic annotations and low-level content descriptors, which allow a board variety of computational studies without the need of obtaining the audio file itself. As described earlier, due to copyright restrictions, the audio data is only provided on request for research purposes or can be obtained by purchasing the set of anthologies which comprise the research corpus. 

Below we first provide an overview of meta-data and automatic annotations which are common to all three subsets and then describe the collections in detail. 

\subsection{Meta-data annotations}
As for the full corpus, we provide for each test collection the editorial meta-information in a machine readable structured text format. The annotations include artist name, style, song title, track duration and the source of the audio file. We furthermore incorporated all tracks in the {\em MusicBrainz} framework. This open online resource holds the editorial meta-data and provides additional information such as artist biographies, user-ratings and links to related tracks and is thus a powerful tool for semantic analysis. For each track, we provide the {\em MusicBrainz ID}, a unique identifier which links the audio file to the corresponding encyclopaedia entry. 

\subsection{Automatic annotations}
For each of the three collections, we provide a number of low-level audio content descriptors on a frame level. Based on such features, a large variety of audio analysis algorithms can be designed and implemented without the need to process the raw audio file itself. The descriptors included in all three collections are listed below and were all extracted in windows of $N=1024$ samples length with $50\%$ overlap (hop size $h=512$) at a sampling rate of $f_s=44.1\mathrm{kHz}$. For stereo signals, both channels were averaged. Further details on the extraction process are provided in the accompanying documentation of the corpus.

\begin{description}
  \item [Spectrum] The magnitudes $|X[k]|$ corresponding to the lower half of the 1024 point Discrete Fourier Transform (DFT) $X[k]$.
  \item [Bark band energies] The spectral energy contained in 28 non-overlapping bands which correspond to an extrapolation of the Bark scale \cite{bark} . 
 \item [MFCCs] The 13 mel-frequency cepstral coefficients (MFCCs) derived from a 40-band filter bank ranging from $0$ to $11\mathrm{kHz}$ \cite{mfccs}.
 \item [Spectral flux] The L2-norm of the spectrum. 
 \item [Spectral rolloff] The frequency in $\mathrm{Hz}$ under which 85\% of the spectral energy is contained. 
 \item [Spectral complexity] The number of peaks present in the local magnitude spectrum in a range between $100\mathrm{Hz}$ and $5\mathrm{kHz}$.
 \item [Spectral flatness] Ratio between geometric and arithmetic mean of the magnitude spectrum in $\mathrm{dB}$. 
 \item [Spectral centroid] The first order central moment of the magnitude spectrum. 
 \item [RMS] Root-mean-square (RMS) of the audio signal.
 \item [ZCR] Zero-crossing rate (ZCR) of the audio signal.
\end{description}

We furthermore provide two automatic annotations related to the melodic content of the singing voice: We extract the predominant melody with the algorithm described in \cite{Melodia}. According to the expected pitch range of flamenco singing, the minimum and maximum frequency were set to $120\mathrm{Hz}$ and $720 \mathrm{Hz}$, respectively. The voicing tolerance was set to $v=0.2$ as suggested in \cite{PolyTrans}, in order to reduce the amount of contour segments corresponding to the guitar accompaniment. Consequently, the obtained pitch contour can be seen as an estimate of the pitch trajectory of the singing voice melody. The analysis window was set to $N=1024$ samples with a hop size of $h=128$ samples at a sampling rate of  $f_s=44.1\mathrm{kHz}$. We furthermore provide automatic note-level transcriptions of the singing voice melody obtained with the system described in \cite{AT}. The transcriptions contain an onset time, duration and MIDI pitch value for each transcribed note and are provided as text and MIDI files. 

\subsection{Test sets}
\subsubsection{cante2midi}
Given the absence of scores in flamenco singing, studies targeting singing voice characteristics often rely on labour-intensive and often to a large extend subjective manual transcriptions. Consequently, automatic and computer-assisted transcription of flamenco singing has become a main objective in the scope of computational flamenco analysis. For the purpose of evaluating automatic singing transcription algorithms, we created a dataset containing 20 tracks taken from the corpus. The collection contains approximately 1 hour and 6 minutes of audio, covering a variety of singers, styles and complexity regarding melodic ornamentation. 

For each track, we provide a manual note-level transcription of the singing voice melody in the standard MIDI format: Each note is defined by its onset time, duration and a semi-tone quantised pitch value. The annotation process was conducted by a person with formal music education and basic knowledge of flamenco and later verified and corrected by a flamenco expert. The output of the transcription system described in \cite{PolyTrans} was taken as a starting point during the annotation process. The annotator manually corrected the transcriptions in a digital audio workstation while listening to both, the original audio track and the transcription synthesised with a piano sound. A possibility was given to mute one of the tracks when necessary. The tuning of the synthesiser was adjusted manually to match the tuning of the audio track. A visual representation of the pitch contour and the baseline transcription was provided as additional aid. In this manner, a total of 6025 ground truth notes were transcribed. 

\begin{figure}[!ht]
 \centerline{
 \includegraphics[width=15.0cm]{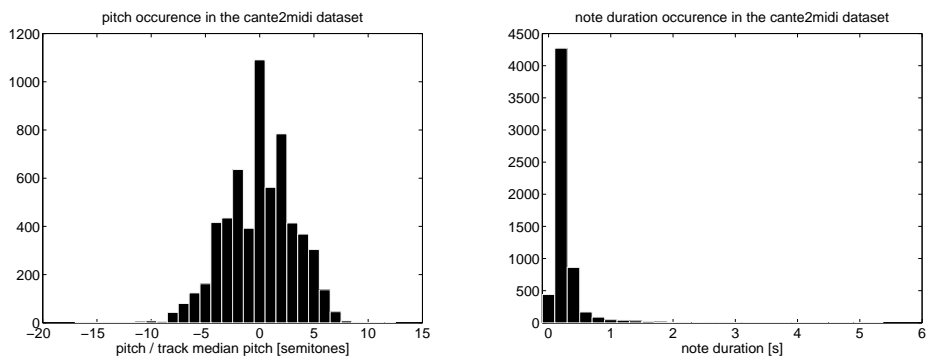}}
 \caption{Relative pitch and note duration occurrences in the {\em cante2midi} dataset.}
 \label{fig:C2Mstats}
\end{figure}

Apart from the ground truth transcriptions, we provide the automatic annotations and meta-data as described above. The systematic description according to \cite{Peeters12} is given in Table \ref{tab:cante2midi} and a short summary of the annotated ground truth data is given in Table \ref{tab:C2Mstats} and Figure \ref{fig:C2Mstats}.   

\begin{table}[t]
\tbl{Overview of the {\em cante2midi} dataset.}{%
\begin{tabular}{| l |c |}
  \hline
 number of tracks & 20 \\
 clip type & full track \\
 number of singers & 15 \\
 total duration & 1h 6m \\
 average track duration & 3m 17s \\
 number of ground truth notes & 6025\\
 average note duration & 0.2640s \\
 note duration standard deviation & 0.2987s \\
 percentage of vocal frames & 42\% \\
   \hline
 \end{tabular}
}
\begin{tabnote}
{\em cante2midi} contains 20 tracks with manually annotated ground truth transcriptions of to the singing voice melody.
\end{tabnote}
\label{tab:C2Mstats}
\end{table}

\subsubsection{canteFAN}
\label{ssec:canteFAN}
The study of characteristic melodic patterns and their repeated occurrence throughout a performance has been of particular interest in the computational analysis of flamenco singing (\cite{pattern}, \cite{pattern2}). Many flamenco styles have evolved from folk music chants and still contain characteristic note sequences which are repeated throughout a song. Flamenco experts can identify not only the style family, but even distinguish the sub style based solely on such a melodic signature. 

An example of a style where repetition of melodic patterns plays a particularly important role is the {\em fandango}: Considered one of the fundamental styles of flamenco, {\em fandangos} have a common formal and harmonic structure. A repeating guitar section, which represents the chorus, alternates with sung verses, which exhibit characteristic reoccurring melodic patterns. Interpretations of {\em fandangos} largely vary with respect to their abstraction from the folkloric origin: While some performances follow a strict rhythm and show only minor modifications of the underlying melodic skeleton, others are characterised by greater rhythmic fluctuations and strong ornamentations, prolongations and variations of the melodic patterns. The discovery and analysis of repeated melodic patterns in {\em fandangos} is consequently of interest for the study of style evolution and intra-style characterisation. 

\begin{figure}[!ht]
 \centerline{
 \includegraphics[width=14.0cm]{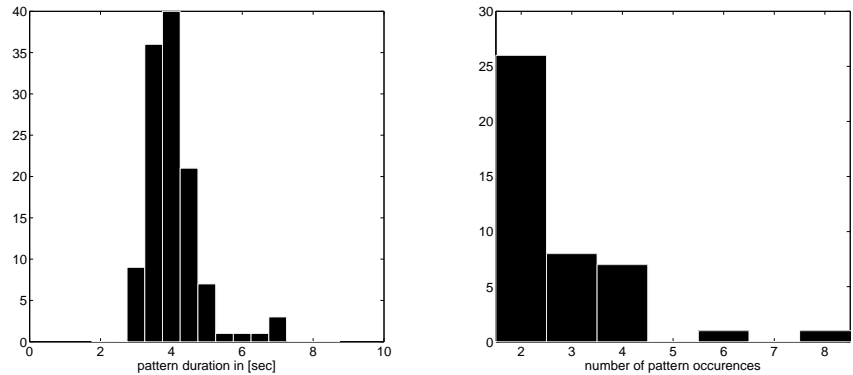}}
 \caption{Pattern duration and occurrences in the {\em canteFAN} dataset.}
 \label{fig:canteFANstats}
\end{figure}

For the purpose of evaluating computational approaches to melodic pattern discovery for flamenco singing we created the {\em canteFAN} dataset: We selected 10 {\em fandango} tracks from the corpus which exhibit a number of reoccurring melodic patterns. In the context of this particular task, we defined a repeated melodic pattern as a small musical unit corresponding to a sung phrase, which is repeated at least once throughout the track. Repetitions can contain minor melodic or rhythmic variations, such as additional or modified grace notes, an overall increase or decrease in tempo or a variation in accentuation.  We manually annotated such patterns and their repetitions in each track in correspondence with flamenco experts. The annotated dataset is available together with the automatic and meta data annotations described above. Table \ref{tab:canteFANstats} and Figure \ref{fig:canteFANstats} give an overview of the dataset and a systematic description is given in Table \ref{tab:canteFAN}.

\begin{table}[t]
\tbl{Overview of the {\em canteFAN} dataset.}{%
\begin{tabular}{| l |c |}
  \hline
 number of tracks & 10 \\
 clip type & full track \\
 number of singers & 6 \\
 total duration & 28m \\
 average track duration & 2m 45s \\
 total number of patterns & 43 \\
 total number of occurrences & 119\\
 average number of pattern occurrences & 2.77\\
 average pattern duration & 4.07s \\
 pattern duration standard deviation & 0.76s \\
   \hline
 \end{tabular}
}
\begin{tabnote}
{\em canteFAN} contains 10 tracks with manual annotations of repeated melodic patterns. 
\end{tabnote}
\label{tab:canteFANstats}
\end{table}

\subsubsection{cante100}
With the purpose of exploring melodic and rhythmic features in flamenco music and in particular their differences across styles, we selected a subset of 100 tracks from the corpus and manually annotated their style family. Applying the same design criteria as for the entire corpus, this subset gives a representative sample of flamenco music with uniform sampling regarding styles. The collection contains a total of 5 hours and 56 minutes of audio recordings and includes 47 singers. While there are numerous styles and sub-styles defined in flamenco \cite{gamboa05}, a standard taxonomy has so far not been established. In correspondence with experts in the field we defined ten style families used in the scope of this data collection: \textit{Tangos y tientos}, \textit{soleares}, \textit{seguiriyas}, \textit{canti{\~n}as}, \textit{buler{\'i}as}, \textit{malague{\~n}as y grana{\'i}nas}, \textit{fandangos}, \textit{cantes mineros}, \textit{ton{\'a}s} and \textit{cantes de ida y vuelta}. A detailed explanation of this categorisation and the included sub-styles is provided in an explanatory document included in the dataset. As depicted in Figure \ref{fig:pie}, these categories cover 96\% of the recordings contained in the full research corpus. Tracks are equally distributed among the defined classes, resulting in 10 tracks per annotated style. 

\begin{figure}[!ht]
 \centerline{
 \includegraphics[width=6.0cm]{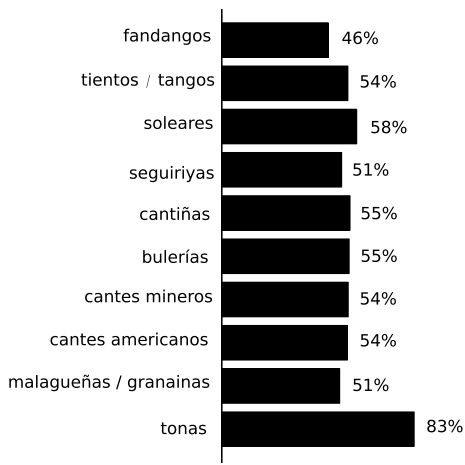}}
 \caption{Vocal sections in the {\em cante100} by style.}
 \label{fig:vocsec}
\end{figure}

In addition, we manually annotated the sections of the song where vocals are present. The task of vocal detection is fundamental to a number of MIR systems targeting the singing voice and consequently there is a need for such ground truth annotations. Figure \ref{fig:vocsec} shows that the percentage of frames in which the vocals are present is mainly consistent throughout the style families. The {\em tonas} present an exception. In these a cappella songs, due to the absence of the guitar, the vocals are present throughout the song except for some short vocal rests.  

A systematic description of the dataset according to \cite{Peeters12} is given in Table \ref{tab:cante100}. Similar to the previously presented datasets, we provide meta data together with automatic annotations. An overview of the database statistics for {\em cante100} is given in table \ref{tab:cante100stats}. 

\begin{table}[t]
\tbl{Overview of the {\em cante100} dataset.}{%
\begin{tabular}{| l |c |}
  \hline
number of tracks & 100 \\
clip type & full track \\
number of included style families & 10\\
number of singers & 48 \\
total duration & approx. 5h 58m \\
average track duration & 3m 35s \\
percentage of vocal frames & 55.17\%\\
   \hline
 \end{tabular}
}
\begin{tabnote}
{\em cante100} contains 100 tracks with manual annotations of style family and vocal sections. 
\end{tabnote}
\label{tab:cante100stats}
\end{table}

\section{Case studies}
Subsequently, we present a number on example applications of computational approaches to flamenco analysis. We evaluate existing methods for vocal detection, automatic singing transcription, detection of repeated melodic patterns and melodic similarity on the previously introduced test collections. We furthermore showcase two explorative data-driven studies targeting the rhythm and tonality across styles.

\subsection{Vocal detection}
In flamenco music, the singing voice represents the central musical element. Consequently, studies mostly focus on analysing the vocals. Therefore, for computational methods, a reliable vocal segment detection is fundamental for a number of analysis algorithms. Related work outside the scope of flamenco singing has addressed the detection of singing voice segments mainly as a machine learning task (\cite{VD1}, \cite{VD2} and \cite{VD3}). While such methods give convincing results, they nevertheless require a large amount of annotated ground truth data and a computationally expensive training phase. 

In the context of flamenco music, related approaches have exploited two key characteristics: The perceptual dominance of the voice with respect to the accompaniment and the limited instrumentation containing mainly vocals and guitar. As a pre-processing stage to a note-level transcription algorithm, \cite{PolyTrans} use a predominant melody extraction algorithm which estimates the pitch contour related to the perceptually dominant sound source. While this assumption holds for large parts of flamenco recordings, the guitar may take over the main melodic line during the introduction or instrumental interludes. Consequently, the authors report mistakenly transcribed guitar contours as a main source of error. Based on these findings, \cite{AT} apply an additional contour filtering stage in order to eliminate contour sections which originate from the guitar accompaniment. 

We evaluated both algorithms for the manually annotated vocal sections in the {\em cante100} dataset. {\em KG-15} denotes the algorithm described in \cite{AT} and {\em PM-raw} the approach described in \cite{PolyTrans}. Both methods were evaluated in frames of length $N=128$ samples at a sample rate of $f_s=\mathrm{kHz}$ by means of {\em voicing precision}, {\em voicing recall} and {\em voicing f-measure}. {\em Voicing precision} is defined as the fraction of all frames estimated as voiced, which are labelled as voiced in the ground truth. {\em voicing recall} corresponds to the fraction of all voiced ground truth frames, which are estimated as voiced. The resulting f-measure is calculated as the harmonic mean of precision and recall. 

\begin{table}[t]
\tbl{Vocal detection evaluation.}{%
 \begin{tabular}{|l|c|c|c|}
  \hline
  & KG-15 &  PM-raw \\
  \hline
  voicing precision & 0.97 & 0.67 \\
  voicing recall  & 0.75 & 0.80 \\
  voicing f-measure  & 0.85 & 0.73 \\
  \hline
 \end{tabular}}
 \label{tab:ATeval}
\begin{tabnote}
Precision, recall and f-measure for two vocal detection schemes evaluated on the {\em cante100} dataset. 
\end{tabnote}
\label{tab:Veval}
\end{table}

The results show that the contour filtering process reduces the number of mistakenly transcribed guitar contours, resulting in an increase in precision. The slightly lower recall indicates that also a small percentage of vocal contours are eliminated. Nevertheless, the f-measure indicates an overall higher performance. 

\subsection{Automatic singing transcription}
Obtaining a note-level transcription from an audio signal is considered on of the most challenging tasks in MIR. \cite{ATMnew} extensively reviewed related approaches and pointed out that generic systems might not cover the characteristics of a specific instrumentation or music tradition. Flamenco singing poses a particular challenge, given the non-percussive and pitch-continuous nature of the singing voice as well as complex melodic progressions and ornamentations and tuning inaccuracies characteristic to flamenco singing. 

A first system proposed by \cite{MonoTrans} for the specific case of a cappella flamenco singing has been extended for accompanied flamenco singing by \cite{PolyTrans}. Recently, a novel transcription system for accompanied flamenco singing was proposed in \cite{AT}. Without going into the algorithm details, we show the evaluation of both transcription systems on the \textit{cante2midi} data collection and compare to the results reported for the monophonic dataset used in \cite{MonoTrans}. The evaluation is carried out in accordance with the measures proposed by the authors of \cite{MonoTrans}: A note is correctly detected, if the onset is located within a tolerance of 15ms, the duration is estimated within a tolerance of 30\% of the ground truth duration and the quantised MIDI pitch is correctly detected. Consequently, the following measures can be defined: 

\begin{description}
  \item [Note precision] Proportion of all detected notes, which are correctly transcribed ground truth notes.
  \item [Note recall] Proportion of all ground truth notes, which are correctly transcribed. 
  \item [Note f-measure] $\frac{2 \cdot precision \cdot recall} {precision+recall}$
\end{description}

\begin{figure}[!ht]
 \centerline{
 \includegraphics[width=9.0cm]{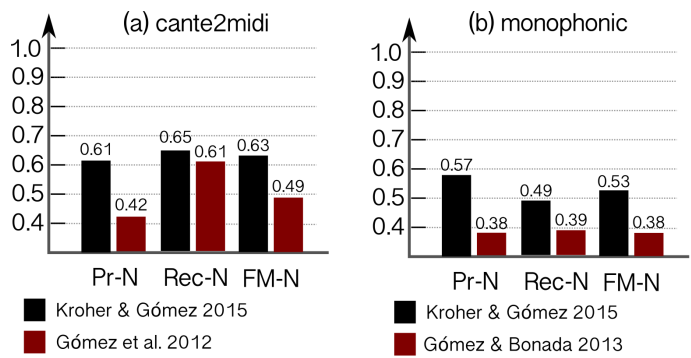}}
 \caption{Evaluation of automatic singing transcription algorithms on the {\em cante2midi} dataset.}
 \label{fig:AST_eval}
\end{figure}

The results displayed in Figure \ref{fig:AST_eval} indicate a better for both methods a better performance on the {\em cante2midi} dataset when compared to the a cappella singing dataset \cite{MonoTrans}. This can be explained with the particular characteristics of a cappella singing styles: Melodies are mainly composed of conjunct degrees and contain a higher amount of melismatic ornamentation, resulting in a more complex note segmentation task. In addition, given the absence of guitar accompaniment, tuning tends to fluctuate during a song. We can furthermore observe an overall better performance of the method proposed in \cite{AT}. 

Apart from automatic vocal melody transcription, further possible applications of the \textit{cante2midi} subset include vocal detection, vocal pitch extraction and studies targeting characteristics of melodic ornamentation. 

\subsection{Inter- and intra-style analysis}
In flamenco music, a particular style is characterised by distinct melodic, rhythmic, and structural features. Consequently, automatic style discrimination is a non-trivial task and we identify a need to evaluate and adapt existing MIR techniques to characterise the particular musical facets inherit to the diverse style categories. Below, we present two data-driven studies in which we analyse audio descriptors related to tonality and tempo across the ten style families contained in {\em cante100} dataset.

\subsubsection{Tonality}
\begin{figure}[!ht]
 \centerline{
 \includegraphics[width=6.5cm]{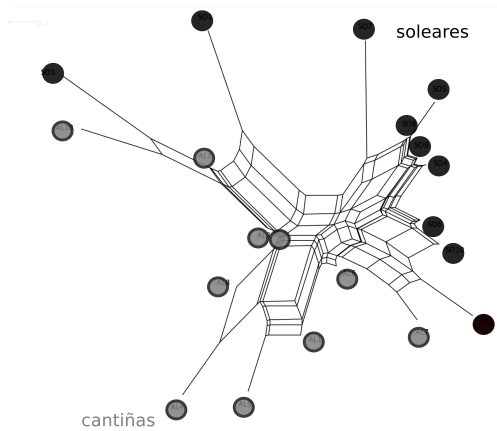}}
 \caption{Phylogenetic tree for pitch histogram distances among \textit{soleares} and \textit{canti{\~n}as}.}
 \label{fig:example1}
\end{figure}

\begin{figure}[!ht]
 \centerline{
 \includegraphics[width=6.5cm]{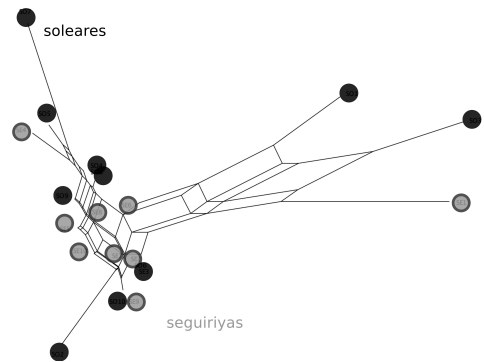}}
 \caption{Phylogenetic tree for pitch histogram distances among \textit{soleares} and \textit{seguiriyas}.}
 \label{fig:example2}
\end{figure}

In a first case study, we evaluate the suitability of statistical melody analysis for the characterisation of style-specific features with respect to expert knowledge: We first calculated pitch histograms from automatic transcriptions for all excerpts in the \textit{cante100} dataset. After shifting each histogram to the most occurring pitch class, assuming it to be the tonic, we compute pair-wise correlations among the histograms and generate phylogenetic trees displaying the distances among the examples in a two-dimensional space. Three examples are displayed in Figures \ref{fig:example1}, \ref{fig:example2} and \ref{fig:example3}. 

When analysing the similarity with respect to statistical note occurrence among \textit{canti{\~n}as} and \textit{soleares} (Figure \ref{fig:example1}), we can observe a separation of the two styles. This coincides with fact that these two styles differ in their underlying tonality: While \textit{soleares} are based on the phrygian mode, \textit{canti{\~n}as} are sung in major mode. In contrast, when comparing \textit{soleares} and \textit{seguiriyas} (Figure \ref{fig:example2}), we do not observe a separation of the styles, since both are sung in phrygian mode. When performing an intra-style analysis of the \textit{buler{\'i}as} style (Figure \ref{fig:example3}), we observe a small cluster of three examples. A listening analysis confirms, that the examples within the observed cluster contain a melody which is strongly centred around the interval structure characteristic to the \textit{andalusian cadence} \cite{gamboa05}, which is not the case for the other analysed tracks. 

This example furthermore indicates the limitations of statistical note analysis for inter- and intra-style discrimination: The underlying tonality is not a definite criteria for discriminating styles, since various styles are based on the same mode. Furthermore, sub-styles often differ in only short melodic sequences, which require a more in depth analysis of the melodic contour, i.e. as described for a cappella styles in \cite{similarity}. 

\begin{figure}[!ht]
 \centerline{
 \includegraphics[width=6.5cm]{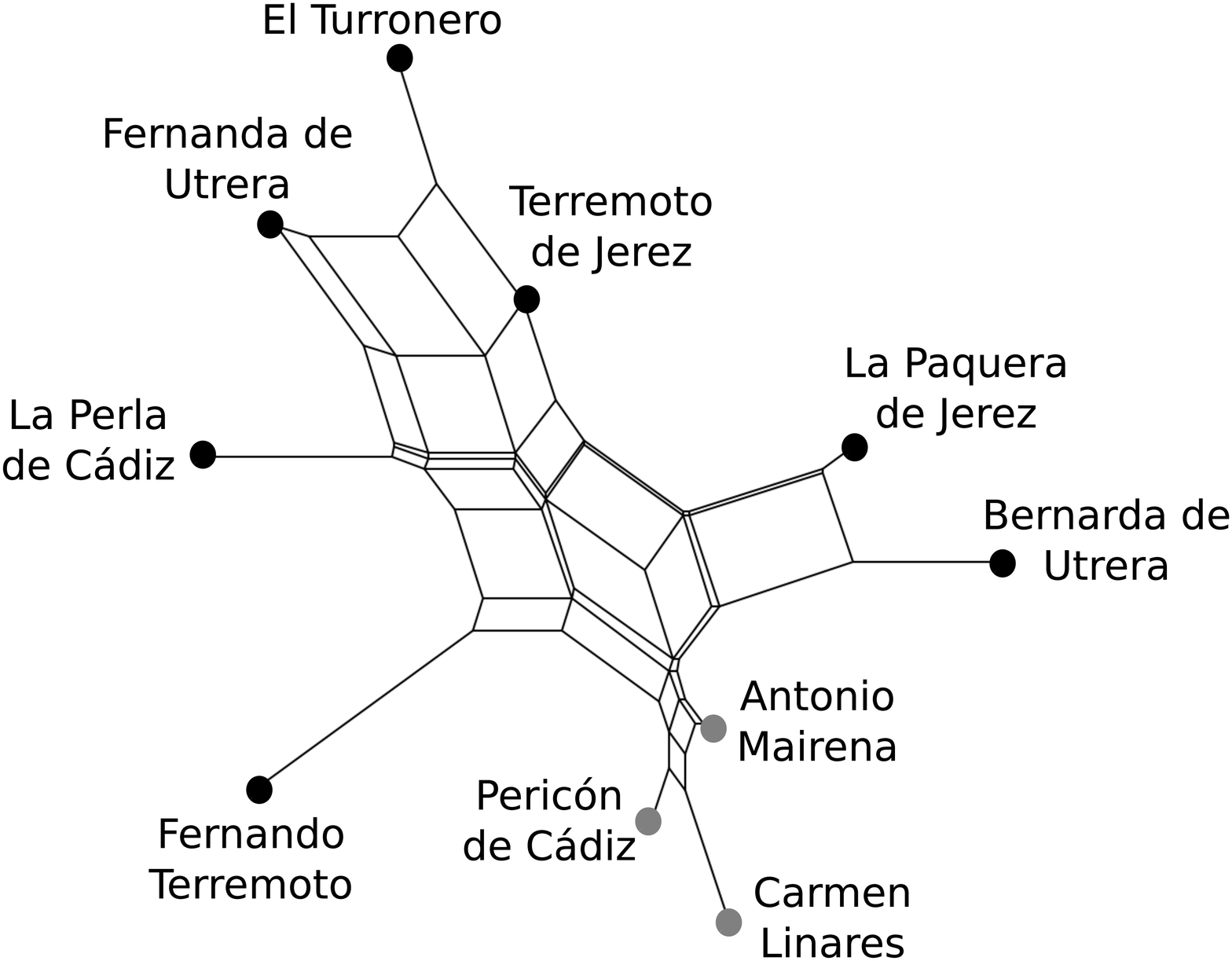}}
 \caption{Phylogenetic tree for pitch histogram distances among \textit{buler{\'i}as}.}
 \label{fig:example3}
\end{figure}

\subsubsection{Tempo}
In an explorative approach, we investigate automatic tempo annotations across the ten style families included in the {\em cante100} dataset. So far, MIR algorithms related to rhythm and tempo have not been evaluated in the context of flamenco music. While many other genres and, to a large extend, Western popular music is characterised by a periodic succession of strong and weak accentuations which follow the underlying rhythm, flamenco music contains more complex and alternating rhythmic structures as well as strong tempo fluctuations.

In this preliminary study, we use the multi-feature beat tracker algorithm described in \cite{beat} to extract two global descriptors for each song: The estimated tempo of the track in beats per minute (BPM) and a confidence value ranging from 0 to 5.32 related to the quality of the corresponding tempo estimate (\cite{beat2}). With the aim of obtaining a compact representation indicating differences among styles, we compute the histogram of estimated BPM values for each style separately. In order to incorporate the beat estimation quality, we weight each histogram contribution with its corresponding confidence factor. In this way, high confidence estimations contribute stronger to the statistic representation than weaker estimates. The sum of histogram bins furthermore serves as an indicator for the difficulty of tempo estimation in the context of a particular style: The larger the sum, the higher the overall tempo estimate quality. 

\begin{figure}[!ht]
 \centerline{
 \includegraphics[width=18cm]{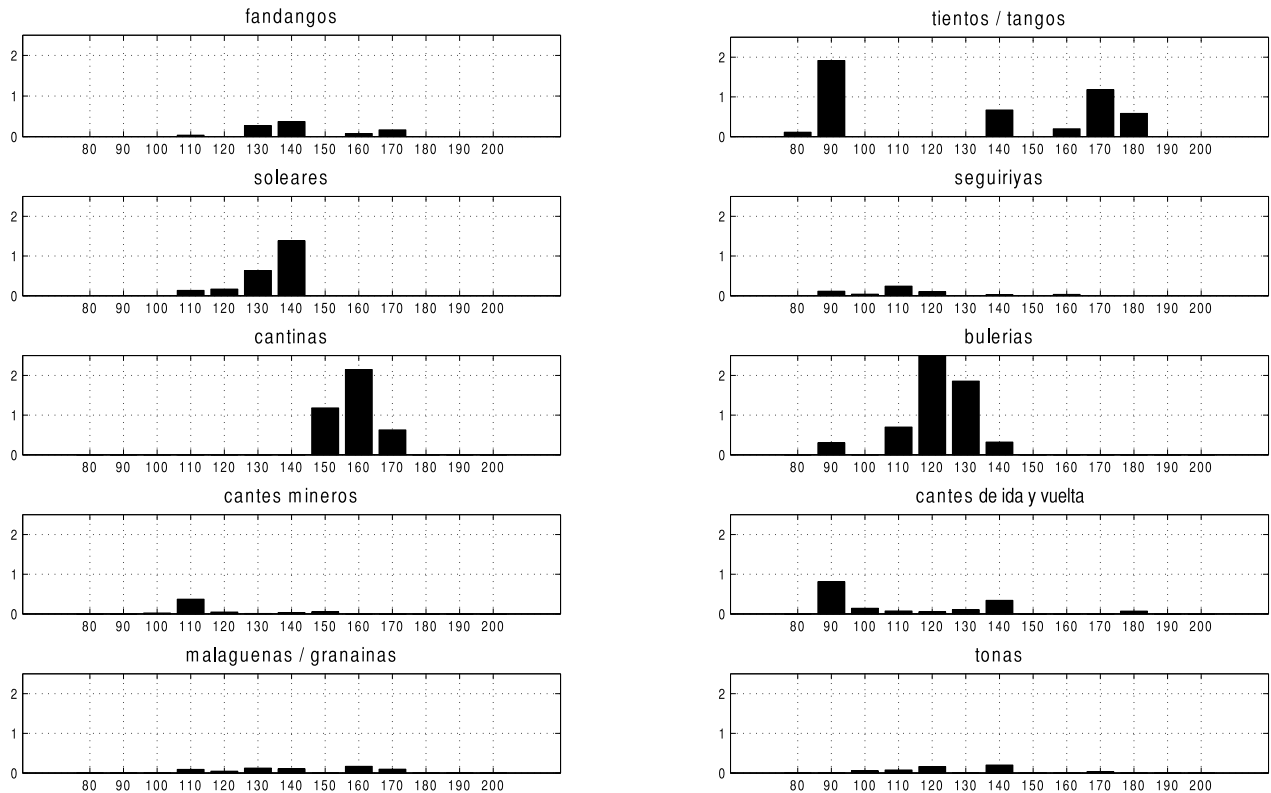}}
 \caption{Weighted tempo histograms by style.}
 \label{fig:example3}
\end{figure}

The resulting histograms for the ten styles under study (Figure \ref{fig:example3}) provide a number of interesting observations: First, there are significant differences among the overall tempo estimation confidences among the analysed styles. While i.e. {\em buler{\'i}as} and {\em canti{\~n}as} give high confidence values, the estimates for {\em seguiriyas} and {\em tonas} appear to be less reliable. These results coincide with the observation that in {\em buler{\'i}as} and {\em canti{\~n}as} the beat is strongly accentuated and often additionally emphasised by hand-clapping. Furthermore, the tempo tends to be more stable compared to other styles. In the histograms with an overall high confidence we can furthermore observe style-specific tempo differences: While the estimates for the {\em buler{\'i}as} are between $90$ and $140$ BPM, the faster {\em canti{\~n}as} give estimates between 150 and 170 BPM. In the family of {\em tientos and tangos} we can observe both, slow (around 90 BPM) and fast (140-180 BPM) examples. This can be explained by the fact that this style family is comprised by two sub-styles which share common melodic and harmonic elements but differ in tempo: The rather slow {\em tientos} and the faster {\em tangos}. 

\subsection{Discovery of repeated melodic patterns}
As described in \ref{ssec:canteFAN}, the automatic discovery of repeated melodic sequences in flamenco recordings is not only fundamental to a variety musicological studies but also provides crucial information for automatic indexing applications. For Western music, prior approaches have mainly used the score to identify melody repetitions. For a complete overview of symbolic approaches we refer to\cite{symbolicPattern}. However, it was reported in \cite{audioPattern}, that applying score based algorithms to automatic transcriptions results in a significant decrease in performance. Therefore, given the absence of scores, an audio-based approach was proposed for flamenco music in \cite{pattern2}. 

The evaluation dataset used in this work included 11 recordings of performances of the {\em fandango} style, most of which are sung with a low degree of ornamentation and variation of the characteristic melodic patterns. The {\em canteFAN} dataset on the other hand includes examples with a varying degree of abstraction in a sense of ornamentation and variation. In Table \ref{tab:Peval} we present the results for both music collections, the {\em canteFAN} dataset and the set of audio examples used by the authors denoted as P-15. The displayed evaluation measures are taken from the related task entitled "Discovery of Repeated Themes and Sections" in the MIREX evaluation framework (\cite{MIREX}): The {\em establishment} measures evaluate how well a repeated pattern is detected by the algorithm, regardless of how well all repetitions have been detected. The  {\em occurrence} measures refer to the capability of retrieving all repetitions of a pattern. For a complete description of evaluation methodology we refer to \cite{MIREX} and \cite{audioPattern}. 

\begin{table}[t]
\tbl{Pattern detection evaluation.}{%
 \begin{tabular}{|l|c|c|}
 \hline
 & canteFAN & P-15\\
  \hline
  establishment precision & 0.47 & 0.48\\
  establishment recall  & 0.61 & 0.78\\
  establishment f-measure  & 0.53 & 0.60\\
  occurrence precision & 0.26 & 0.23\\
  occurrence recall  & 0.37 & 0.56\\
  occurrence f-measure  & 0.31 & 0.33\\
  \hline
 \end{tabular}}
 \label{tab:Peval}
\begin{tabnote}
Obtained results for establishment and occurrence measures in the task of detecting repeated melodic patterns. 
\end{tabnote}
\label{tab:Peval}
\end{table}

While the algorithm shows a similar behaviour for the establishment and occurrence precision for both datasets, the recall is significantly lower for the {\em canteFAN} collection, resulting in an overall lower f-measure. Consequently, more patterns and their repetitions remain undiscovered by the algorithm. This might be related to the fact that the examples contained in {\em canteFAN} contain a larger amount of melodic ornamentation and variation. 

\section{Conclusions and future work}
We presented \textit{corpusCOFLA}, a research corpus for the computational analysis of flamenco singing. We explained the design criteria, justified the selection of audio examples regarding established paradigms and pointed out the particular challenges of a research corpus creation for flamenco music. We furthermore described three test collections drawn from the corpus and gave examples for possible applications in the context of music information retrieval. We have several goals for the future development and augmentation of the corpus: The editorial meta data will be completed with a systematic style taxonomy in order to facilitate style and sub-style specific queries. We furthermore plan to include further information, such as lyrics, guitarist or year of production, to make this corpus suitable for a larger variety of computational studies. Finally, we aim to create further test collections for specific music information retrieval tasks and their application to flamenco music, such as inter- and intra-style similarity, performance analysis and cover song identification.


\bibliographystyle{ACM-Reference-Format-Journals}
\bibliography{acmlarge-sample-bibfile}


\elecappendix
\section{Systematic description of the test collections}

\begin{table}[t]
\tbl{\textit{cante2midi} }{%
\begin{tabular}{p{8cm}}
 \hline
 \textbf{(C1) Corpus ID:} corpus:MIR:COFLA:cante2midi:2015:ver1.0\\
 \hline
 \textbf{(A) Raw Corpus}\\
 \textbf{(A1) Definition:} (a13) real sampled items; 20 tracks taken from commercially available flamenco anthologies; gathered for the specific purpose of evaluating automatic transcription systems of strongly ornamented singing; uniform sampling regarding degree and complexity of melodic ornamentation; \\
 \textbf{(A2) Type of media diffusion:} References to audio sources (commercially available CDs) with CD and track number; Audio sources are shared on request for research purposes only; Manual and automatic annotations publicly available from online repository. \\
 \hline
 \textbf{(B) Annotations}\\
 \textbf{(B1) Origin:} (b11) synthetic and (b15) manual; \\
 \textbf{(B21) Concepts definition:} Automatic vocal melody transcription;\\
 \textbf{(B22) Annotation rules:} (b15) note-level representation of the vocal melody, editorial meta-data and {\em musicBrainz} ID; (b11) predominant melody (\cite{Melodia}); automatic note-level transcriptions (\cite{AT}); Frame-wise low-level audio descriptors (\cite{essentia}): Spectrum, bark band energies, MFCCs, spectral flux, spectral rolloff, spectral complexity, spectral flatness, spectral centroid, RMS, zero-crossing rate. \\
 \textbf{(B31) Annotators:} (b15) person with formal music education, experience in melody transcription and basic knowledge of flamenco.\\
 \textbf{(B32) Validation:} correction process in correspondence with flamenco experts. \\
 \textbf{(B4) Annotation tools:} (b15) MIDI editor (Logic Pro X), visual f0 representation (Matlab); (b11) Vamp plugin {\em Melodia}, binary implementation of \cite{AT}, {\em essentia} c++ audio analysis library; \\
 \hline
 \textbf{(C) Documents and Storing}\\
 \textbf{(C2) Storage of the Created Annotations:} (b11) standard MIDI files (.mid), text files containing note event information (.csv) (b11) standard MIDI files (.mid), text files containing note event information, text files containing note event information (.csv), text files containing predominant melody (.csv), text files containing low-level descriptors (.csv), structured text files (.xml) containing meta-data and {\em musicBrainz} ID.\\
 \hline

 \end{tabular}}
\begin{tabnote}
Systematic description of the \textit{cante2midi} collection.
\end{tabnote}
\label{tab:cante2midi}
\end{table}

\begin{table}[t]
\tbl{\textit{canteFAN}}{%
\begin{tabular}{p{8cm}}
  \hline
 \textbf{(C1) Corpus ID:} corpus:MIR:COFLA:canteFAN:2015:ver1.0\\
 \hline
 \textbf{(A) Raw Corpus}\\
 \textbf{(A1) Definition:} (a13) real sampled items; 10 tracks taken from commercially available flamenco anthologies belonging to the {\em fandango} style; gathered for the specific purpose of evaluating algorithms for the discovery of repeated melodic patterns; uniform sampling regarding degree and complexity of ornamentation and variation of the characteristic patterns; \\
 \textbf{(A2) Type of media diffusion:} References to audio sources (commercially available CDs) with CD and track number; Audio sources are shared on request for research purposes only; Manual and automatic annotations publicly available from online repository. \\
 \hline
 \textbf{(B) Annotations}\\
 \textbf{(B1) Origin:} (b11) synthetic and (b15) manual; \\
 \textbf{(B21) Concepts definition:} Repeated melodic patterns;\\
 \textbf{(B22) Annotation rules:} (b15) start end end times and pattern index of repeated melodic sequences, editorial meta-data and {\em musicBrainz} ID; (b11) predominant melody (\cite{Melodia}); automatic note-level transcriptions (\cite{AT}); Frame-wise low-level audio descriptors (\cite{essentia}): Spectrum, bark band energies, MFCCs, spectral flux, spectral rolloff, spectral complexity, spectral flatness, spectral centroid, RMS, zero-crossing rate. \\
 \textbf{(B31) Annotators:} (b15) person with formal music education and limited knowledge of flamenco music.\\
 \textbf{(B32) Validation:} correction process in correspondence with flamenco experts. \\
 \textbf{(B4) Annotation tools:} (b15) Sonic Visualizer; (b11) Vamp plugin {\em Melodia}, binary implementation of \cite{AT}, {\em essentia} c++ audio analysis library; \\
 \hline
 \textbf{(C) Documents and Storing}\\
 \textbf{(C2) Storage of the Created Annotations:} (b11) text files containing pattern index, start and end times (.txt) (b11) standard MIDI files (.mid), text files containing note event information, text files containing note event information (.csv), text files containing predominant melody (.csv), text files containing low-level descriptors (.csv), structured text files (.xml) containing meta-data and {\em musicBrainz} ID.\\
 \hline
 \end{tabular}
}
\begin{tabnote}
Systematic description of the \textit{canteFAN} collection.
\end{tabnote}
\label{tab:canteFAN}
\end{table}
\begin{table}[t]
\tbl{\textit{cante100}}{%
\begin{tabular}{p{8cm}}
   \hline
 \textbf{(C1) Corpus ID:} corpus:MIR:COFLA:cante100:2015:ver1.0\\
 \hline
 \textbf{(A) Raw Corpus}\\
 \textbf{(A1) Definition:} (a13) real sampled items; 100 tracks taken from commercially available flamenco anthologies belonging to ten different style families; gathered as a representative subsample of the {\em corpusCOFLA} flamenco research corpus; uniform sampling regarding style: ten example for each of the ten defined style family; \\
 \textbf{(A2) Type of media diffusion:} References to audio sources (commercially available CDs) with CD and track number; Audio sources are shared on request for research purposes only; Manual and automatic annotations publicly available from online repository. \\
 \hline
 \textbf{(B) Annotations}\\
 \textbf{(B1) Origin:} (b11) synthetic and (b15) manual; \\
 \textbf{(B21) Concepts definition:} Style family and vocal sections;\\
 \textbf{(B22) Annotation rules:} (b15) manually annotated style family from a previously defined set of ten categories, frame-wise annotation of vocal segments, editorial meta-data and {\em musicBrainz} ID; (b11) predominant melody (\cite{Melodia}); automatic note-level transcriptions (\cite{AT}); Frame-wise low-level audio descriptors (\cite{essentia}): Spectrum, bark band energies, MFCCs, spectral flux, spectral rolloff, spectral complexity, spectral flatness, spectral centroid, RMS, zero-crossing rate. \\
 \textbf{(B31) Annotators:} (b15) style families: flamenco experts; vocal sections: person with formal music education and limited knowledge of flamenco music.\\
 \textbf{(B32) Validation:} style families: agreement among various flamenco experts. \\
 \textbf{(B4) Annotation tools:} (b15) Sonic Visualizer; (b11) Vamp plugin {\em Melodia}, binary implementation of \cite{AT}, {\em essentia} c++ audio analysis library; \\
 \hline
 \textbf{(C) Documents and Storing}\\
 \textbf{(C2) Storage of the Created Annotations:} (b11) text files containing vocal section annotations (.csv), structured text files (.xml) containing the annotated style family (b11) standard MIDI files (.mid), text files containing note event information, text files containing note event information (.csv), text files containing predominant melody (.csv), text files containing low-level descriptors (.csv), structured text files (.xml) containing meta-data and {\em musicBrainz} ID.\\
 \hline \end{tabular}
}
\begin{tabnote}
Systematic description of the \textit{cante100} collection.
\end{tabnote}
\label{tab:cante100}
\end{table}




\end{document}